\documentclass[journal,showpacs]{IEEEtran}
\usepackage{amsmath,amssymb,graphicx}
\usepackage{multirow}

\ifCLASSINFOpdf
\else
\fi
\begin{document}
%

\title{Quantum Key Distribution with High Order Fibonacci-like Orbital Angular Momentum States}
%
%

\author{Ziwen Pan, Jiarui Cai, Chuan Wang
\thanks{Z. Pan is with State Key Laboratory of Information Photonics and Optical
Communications, School of Information and Communication Engineering, Beijing University of Posts and Telecommunications,
Beijing 100876, China}
\thanks{J. Cai is with State Key Laboratory of Information Photonics and Optical
Communications, School of Information and Communication Engineering, Beijing University of Posts and Telecommunications,
Beijing 100876, China}
\thanks{C. Wang is with the School of Science and the State Key Laboratory of Information Photonics and Optical Communications, Beijing University of Posts and Telecommunications, Beijing 100876, China, e-mail: wangchuan@bupt.edu.cn.}}

\markboth{Journal of \LaTeX\ Class Files,~Vol.~14, No.~18, August~2016}%
{Shell \MakeLowercase{\textit{et al.}}: Bare Demo of IEEEtran.cls for IEEE Journals}


\maketitle

\begin{abstract}
The coding space in quantum communication could be expanded to high-dimensional space by using orbital angular momentum (OAM) states of photons, as both the capacity of the channel and security are enhanced. Here we present a novel approach to realize high-capacity quantum key distribution (QKD) by exploiting OAM states. The innovation of the proposed approach relies on a unique type of entangled-photon source which produces entangled photons with OAM randomly distributed among high order Fiboncci-like numbers and a new physical mechanism for efficiently sharing keys. This combination of entanglement with mathematical properties of high order Fibonacci sequences provides the QKD protocol immunity to photon-number-splitting attacks and allows secure generation of long keys from few photons. Unlike other protocols, reference frame alignment and active modulation of production and detection bases are unnecessary.
\end{abstract}




\section{Introduction}

\IEEEPARstart{S}{ecure} keys generation between distant users could be realized by using quantum key distribution (QKD) \cite{bennett1984quantum,qkd1,qkd2,qkd3}. QKD is to encode and decode the classical keys on the quantum state of single photons, traditionally on the polarization degrees of freedom or the phase degrees of freedom. The non-cloning principle guarantees the security of the QKD process, and that any eavesdropping behavior could be discovered by the communication parties.


Recently there has been much work on QKD \cite{Trojan,Trojan2,AQKD,GKCVQKD,DECOYMDI,XKsong3level}.
However, the noise, the high channel loss and low channel capacity are three main disadvantages of quantum communications. During the past decades, various approaches are exploited to purify the entanglement \cite{EP2,EP3,EP4,EP5,EP6,EP7} and to reduce the channel loss extending the communication distance \cite{repeater1,repeater2,quantumStateAmplification1,quantumStateAmplification2, quantumStateAmplification3,quantumStateAmplification4,quantumStateAmplification5, quantumStateAmplification6}. On the other hand, increasing the dimension of quantum systems would increase the channel capacity which brings several advantages for quantum communication \cite{bechmann2000quantum,r14,highbenefit2013}. For example, the coding capacity and the security are increased along with the increment of the Hilbert space, and with the increment of the mutually unbiased bases \cite{schwinger1960unitary,ivonovic1981geometrical,wootters1989optimal,lawrence2002mutually}. Here the larger Hilbert space could be realized by using multi-level atoms \cite{multi1,multi2}, hyperentangled photons \cite{wang2005quantum,bechmann2000quantum3state}, or the orbital angular momentum (OAM) state of single photons \cite{oemrawsingh2005experimental,nagali2010experimental,leach2010quantum}.

Especially, by using OAM states of photons \cite{allen1992orbital}, several novel applications in quantum information have been proposed \cite{OAM1,OAM2,OAM3}. For instance, golden angle (GA) spiral arrays can be applied to generate multiple OAM values encoding well-defined numerical sequences on their farfield radiation patterns \cite{GA}. It may also be possible to combine GA spirals with spontaneous parametric down conversion (SPDC) in a nonlinear crystal to engineer a new type of entangled light source, which produces photon pairs whose OAM values' summation is a Fibonacci number. This novel setup allows efficient production of states with large OAM values in quantum communications \cite{mirhosseini2013efficient}.

Here in this study, we propose a high-capacity coding method and an efficient QKD scheme by employing high order Fibonacci sequence recursion onto OAM states to create a different optical spiral. The third order Fibonacci recursion provides several properties for us to to improve the security from eavesdropping through generating encryption keys with large numbers of digits by using much smaller numbers of photons.

\section{Quantum key distribution using OAM states encoded by third-order Fibonacci sequence}

Recently, Simon et al. presented a high-capacity QKD scheme by randomly encoding the Fibonacci sequence onto entangled OAM states \cite{fibonacci}. Here by introducing the higher order Fibonacci recursion relation, we can improve the QKD protocol in security and capacity. We start with the third-order recursion relation: "$F_n$ = $F_{n-1} + F_{n-2} + F_{n-3}$". Here we assign the first three values of the sequence as $F_1 = 1, F_2 = 2$ and $F_3 = 3$. The optical vortices which have been employed for communication  can be generated in several different ways: spiral phase plate can generate optical vortices with OAM equals to $l\hbar$;  the transformation from Hermite-Gaussian (HG) mode to Laguerre-Gaussian (LG) mode can also be useful to generate the optical vortices we need. Diffractive optical elements can provide another way to generate the optical vortices, for example, by using inhomogeneous anisotropic media \cite{Papalo}.

\subsection{Basic setup}

 In our scheme, the source light is prepared in a superposition state with OAM values equal to third-order Fibonacci numbers as (1, 2, 3, 6, 11, 20, 37, 68, ...). And we choose $N$ consecutive values, $F$ = \{$F_{n_0}, F_{n_0 + 1}, ... , F_{n_0 +N-1}$\}, and assign a block of binary digits to each so that equal numbers of 0's and 1's occur. Here $n_0$ is an arbitrary sequence number indicating the first Fibonacci OAM value we choose. Each photon should be able to generate $log_2 N$ bits of information if OAM values in this set are used. For simplicity, we set $N = 8$ to illustrate the principle for qubits coding. The third-order Fibonacci numbers from 1 to 68 can be assigned with three-digit blocks as follows: $1$ = $000$, $2=001$, $3=010$, $6=011$, $11=100$, $20=101$, $37=110$, $68=111$. Three key digits could be encoded on the OAM of a single photon. The SPDC spiral bandwidth is able to span the largest gap in $F$. It is obvious that both information capacity and security could be improved with larger sets $F$. To simplify the case, we assume that OAM sorters only allow positive OAM values to reach the detectors.

\begin{figure}
\centering
\includegraphics[width=3in]{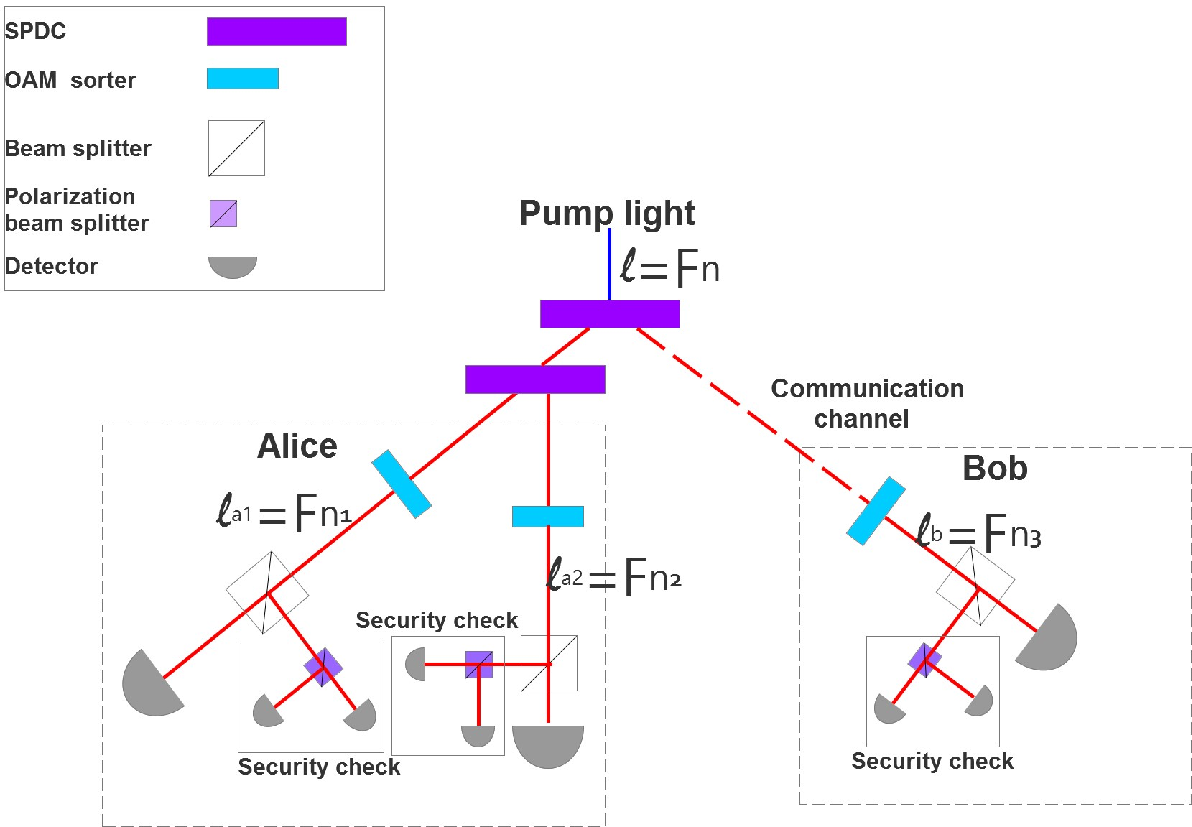}
\caption{Setup for the implementation of QKD with Fibonacci-valued OAM states. The pump light is used to pump the nonlinear crystal, producing signal-idler pairs each with three photons through SPDC process. Alice chooses randomly one photon of each pair to send to Bob and keeps the remaining two photons. The OAM sorters are arranged to only allow photons to reach single-photon detectors if they are third-order Fibonacci-valued, with OAM $l_{a_1}=F_{n_1}$, $l_{a_2}=F_{n_2}$ and $l_b=F_{n_3}$. These OAM values fulfill the relation: $ l = F_n $, $ l =  l_{a_1} + l_{a_2}+ l_b$. Only pairs of values $l_{a_1}$, $l_{a_2}$ and $l_b$ which sum to $F_n$ values are kept.}
\label{f1}
\end{figure}

Please note, as in the rest part of the passage, the subscript $a_1$ and $a_2$ correspond to the two OAM values Alice obtains or the corresponding classic bit depending on the variable and $b$ does the same for Bob. The subscript $n$ indicates the total OAM's sequence number in the Fibonacci sequence before SPDC while $n_1$, $n_2$ and $n_3$ correspond to the OAM after SPDC, as illustrated in Fig.\ref{f1}. These also work for other corresponding variables like classic bits as illustrated in TABLE \ref{t1} .


As shown in Fig.\ref{f1}, two SPDC processes are exploited for the generation of entangled OAM states in third order Fibonacci sequence. After the first SPDC process, one of each pair ($l_b$) is sent straight to Bob while the OAM sorters are arranged to allow only photons with OAM in $F$ ($l_b=F_{n_i}$) to register. The other photon($l_a$), however, needs another SPDC process to be turned into pairs ($l_{a_1}, l_{a_2}$) before sending to Alice. The photons reserved by Alice should also be sent through the same OAM sorters as Bob's to make sure that only photons with OAM equal to third-order Fibonacci numbers are eventually retained. As the collinear SPDC (type \uppercase\expandafter{\romannumeral1} or type \uppercase\expandafter{\romannumeral2}) OAM conservations implies the relations $l_{a_1}+l_{a_2}+l_b=l$. In this case, the third-order Fibonacci recursion forces $l_{a_1}$, $l_{a_2}$ and $l_b$ to be the three adjacent third-order Fibonacci numbers immediately preceding $l=F_n$. For example, by taking $l=F_n =F_6=20$, $l_{a_1}$, $l_{a_2}$ and $l_b$ should be in the set $\{F_{n-1}=F_5=11,F_{n-2}=F_4=6,F_{n-2}=F_3=3\}$.

Moreover, we develop a set of regulations which allows Alice and Bob to generate the three bits' communication through classical channel. And the quantum keys could be generated by the classical calculation on the shared bits.
\begin{table*}[!hbp]
\normalsize
\centering
\begin{tabular}{|p{3cm}|*{11}{p{0.8cm}|}}
\hline
n (sequence number)& 1 & 2 & 3 & 4 & 5 & 6 & 7 & 8 & 9 & 10 \\
\hline
$F_n$ (Fibonacci OAM value)& 1 & 2 & 3 & 6 & 11 & 20 & 37 & 68 & 125 & 230 \\
\hline
$B_n$ (Classical bits/ recognition flag) & 0 & 0 & 0 & 0 & 1 & 1 & 1 & 1 & 0 & 0 \\
\hline
Center(c) or edge(e) flag & e & c & c & e & e & c & c & e & e & c \\
\hline
\end{tabular}
\caption{The bit allocated for each Fibonacci OAM value. Please notice that the bits in this table are not the exact ones used in classical channel. The exact classical bits sending rules are listed in \ref{alice} and \ref{bob}.}
\label{t1}
\end{table*}
As is illustrated in TABLE \ref{t1}, we allocate one binary bit to each OAM value under the rules as: if ($n$ mod $8) \in \{1,2,3,4\}$, then $B_n=0$, otherwise $B_n=1$. Here in TABLE \ref{t1}, we can see that there are only two kinds of bits: $B_1, B_4, B_5, B_8, B_9, ...$ are on the "edge" of each set while $B_2, B_3, B_6, B_7, B_{10}, ...$ are in the "center" of each set.

Based on Alice's possible reception of photons, we can divide all possible situations into 5 categories, as illustrated in TABLE \ref{tbit}. For example, if Alice measures two adjacent OAM values in the third-order Fibonacci sequence, like $|F_6\rangle \otimes  |F_7\rangle$ with $(B_{a_1},B_{a_2})=(1,1)$ in the third row of TABLE \ref{tbit}, then  Bob could have the photon with $l_b=F_5$ ($B_b=B_5=1$) or $F_8$ ($B_b=B_8=1$).
In comparison, the classification from Bob's angle is much simple who has only one photon thus he can only anticipate three possible evaluations for original OAM value $l$.

\begin{table*}[!hbp]
\normalsize
\centering
\begin{tabular}{|c|c|c|c|c|c|c|}
\hline
\multicolumn{4}{|c|}{Alice} & \multicolumn{3}{c|}{Bob}\\
\hline
\multicolumn{2}{|c|}{type} & $(l_{a_1},l_{a_2})$ & $(B_{a_1},B_{a_2})$ & type & $l_b$ & $B_b$ \\
\hline
\multirow{3}{*}{adjacent} & (c,c) & $(F_6,F_7)$ & (1,1)& e & $F_5/F_8$ & 1/1\\
\cline{2-7} & (c,e) & $(F_7,F_8)$ & (1,1) & c/e & $F_6/F_9$ & 1/0\\
\cline{2-7} & (e,e) & $(F_8,F_9)$ & (1,0) & c & $F_7/F_{10}$ & 1/0\\
\hline
\multirow{2}{*}{discontinuous} &\multirow{2}{*}{(c,e)} & $(F_6,F_8)$& (1,1) & c & $F_7$ & 1\\
\cline{3-7} &  & $(F_7,F_9)$ & (1,0) & e & $F_8$ & 1\\
\hline
\end{tabular}
\caption{This table states all five possible situations regarding Alice's reception of photons. Bob's possible reception in every situation is also given. The OAM values given here ($l_{a_1},l_{a_2},l_b$) are only examples to elaborate.}
\label{tbit}
\end{table*}

\subsection{The regulations of the QKD protocol}

Here we present a set of regulation for both Alice and Bob to send binary bits to each other in order to realize practical communication. We use corner marks $n_1$ and $n_2$ to indicate the sequence number of the two Fibonacci OAM values ($l_{a_1}=F_{n_1}$ and $l_{a_2}=F_{n_2}$) Alice obtains and use $n_3$ to do the same for Bob's value ($l_b=F_{n_3}$). Regarding the two SPDC processes, we know that $F_{n_0}\le F_n=F_{n_1}+F_{n_2}+F_{n_3}\le F_{n_0+N-1}$.

As is illustrated in TABLE \ref{alice}, on Alice's side, her first rule is that if she obtains two photons with OAM values equal to $F_{n_1}$ and $F_{n_2}$ where $n_1$ and $n_2$ satisfies:
$ \left\{
\begin{aligned}
&n_2  =  n_1 + 2 \\
&B_{n_1}  =  B_{n_2}= p
\end{aligned}
\right.
$, then Alice sends $!p$ to Bob first. After that Alice will send another $!p$ to Bob after receiving the responding bit from Bob. Otherwise as her second rule, Alice sends either $B_{n_1}$ or $B_{n_2}$ to Bob first and sends $B_{n_2}$ or $B_{n_1}$ to Bob after receiving the responding bit from Bob. (Here by "!" we mean to take the opposite value: $!0=1$, $!1=0$)

\begin{table*}[!hbp]
\normalsize
\centering
\begin{tabular}{|c|c|c|}
\hline

\multirow{2}{*}{Alice ($F_{a_1}=F_{n_1}$, $F_{a_2}=F_{n_2}$)} & The first bit Alice  &The second bit Alice  \\
 & sends to Bob & sends to Bob\\
\hline
$n_2=n_1+2, B_{n_1}=B_{n_2}=p$ & $!p$ & $!p$\\
\hline
 $n_2=n_1+2, B_{n_1}\not=B_{n_2}$& \multirow{2}{*}{$B_{n_1}$($B_{n_2}$)} & \multirow{2}{*}{$B_{n_2}$($B_{n_1}$)} \\
\cline{1-1}
$n_2=n_1+1$ & & \\

\hline
\end{tabular}
\caption{Alice's regulations. }
\label{alice}
\end{table*}

On Bob's side, here we also define two rules for coding, as is illustrated in TABLE \ref{bob}. First we give the definition of a "centeral" (c) bit and an "edge" (e) bit: in TABLE \ref{t1}, the third line gives the classical bit allocated to every OAM value. The bit here comes in the pattern as "0000111100001111...", block after block with four same bits as a block. So we look into every block and we can discover that the second and third one differ from the first and the fourth one when Bob is trying to "guess" Alice's bit based on the first bit Alice sends to him (as is illustrated above, Alice sends her two bits separately). So we simply define the first and the fourth bit in every block as an "edge" (e) block as they are on the edge of its block, and the second and the third as the "centeral" (c) ones, as in TABLE \ref{tbit}. Also they can be described in math: for $F_{n_3}$ or $B_{n_3}$:
$ \left\{
\begin{aligned}
&c(enter): (n_3 \text{ mod } 4)\in\{2, 3\} \\
&e(dge) : (n_3 \text{ mod } 4)\in\{0, 1\}
\end{aligned}
\right.
$.

Now, if Bob's bit is a "central" bit. Then Bob would send his bit (a recognition flag) to tell Alice whether or not he can dope out at least one of her values after she sends her first bit: "$1$" as a "Yes!" and "$0$" can illustrate his inability to dope out any of value.
To describe this situation in math: Bob has a "central" bit $\rightarrow$
$ \left\{
\begin{aligned}
&\text{Alice sends Bob } B_{a_1} \not= B_b = B_{n_3}, \text{ then Bob sends } "1"\\
&\text{Alice sends Bob } B_{a_1} = B_b = B_{n_3}, \text{ then Bob sends } "0"
\end{aligned}
\right.
$.

Second, if Bob has an "edge" bit
$\rightarrow$
$ \left\{
\begin{aligned}
&\text{Alice sends Bob } B_{a_1} \not= B_b = B_{n_3}, \text{ then Bob sends } "1"\\
&\text{Alice sends Bob } B_{a_1} = B_b = B_{n_3}, \text{Bob sends} "!(n_3 \text{ mod }4)"
\end{aligned}
\right.
$.
The first one is the same "recognition flag" as above while the second one would illustrate if Bob's bit is on the "right" edge or the "left" edge of its block.

\begin{table*}[!hbp]
\normalsize
\centering
\begin{tabular}{|c|c|c|c|}
\hline
Bob's bit $B_b$ (c/e)  & The first bit Bob &\multirow{2}{*}{The bit Bob sends to Alice} \\
($F_{b}=F_{n_3}$) &receives from Alice($B_{a_1}$)& \\
\hline
\multirow{2}{*}{c} & $B_{a_1}\not=B_b$ & 1\\
\cline{2-3}
 & $B_{a_1}=B_b$ & 0\\
\hline
\multirow{2}{*}{e} & $B_{a_1}\not=B_b$ & 1\\
\cline{2-3}
 & $B_{a_1}=B_b$ & $!(n_3 \text{ mod }4)$\\
\hline
\end{tabular}
\caption{Bob's regulations. }
\label{bob}
\end{table*}

The whole regulation is showed in TABLE \ref{ab}.

\begin{table*}[!hbp]
\normalsize
\centering
\begin{tabular}{|c|c|c|c|c|}
\hline
Alice($n_1,n_2$) & Bob($n_3$) & Alice sends to Bob & Bob sends to Alice & Alice sends to Bob \\

\hline
\multirow{4}{*}{$n_2=n_1+2, B_{n_1}=B_{n_2}=p$} & \multirow{2}{*}{c} & $!p\not=B_b$ & 1 & $!p$ \\
\cline{3-5}
 & & $!p=B_b$ & 0 & $!p$ \\
\cline{2-5}
&\multirow{2}{*}{e} & $!p\not=B_b$ & 1 & $!p$ \\
\cline{3-5}
 & & $!p=B_b$ & $!(n_3\text{ mod }4)$ & $!p$ \\
\hline
$n_2=n_1+2, B_{n_1}\not=B_{n_2}$ & \multirow{2}{*}{c} & $B_{n_1}(B_{n_2})\not=B_b$ & 1 & $B_{n_2}(B_{n_1})$ \\
\cline{3-5}
\multirow{2}{*}{or}& & $B_{n_1}(B_{n_2})=B_b$ & 0 & $B_{n_2}(B_{n_1})$ \\
\cline{2-5}
 & \multirow{2}{*}{e} & $B_{n_1}(B_{n_2})\not=B_b$ & 1 & $B_{n_2}(B_{n_1})$ \\
\cline{3-5}
$n_2=n_1+1$ & & $B_{n_1}(B_{n_2})=B_b$ & $!(n_3\text{ mod }4)$ & $B_{n_2}(B_{n_1})$ \\
\hline
\end{tabular}
\caption{The whole regulations. }
\label{ab}
\end{table*}


\subsection{The procedures of QKD}

Now we analyse the procedures of QKD in detail. First we assume $x = (-1)^{3-(n_3 \text{ mod }4)}$ to simplify the discussion below and express the exchange process from Bob's perspective.

Bob has the photon with OAM value equal to $F_{n_3}$ where $(n_3$ mod $4)\in\{2, 3\}$(Bob's bit is a central bit).
\begin{enumerate}
\item
If Alice sends Bob $B_{a_1} \not= B_b = B_{n_3}$, then it will be easy for Bob to know exactly which OAM value Alice has.
\begin{enumerate}
\item
Alice sends her bit according to her second rule.

This means that Alice has $|F_{n_3+x}\rangle\otimes |F_{n_3+2*x}\rangle$, so she would surmise that Bob's photon  $|F_b\rangle =|F_{n_3}\rangle$ or $|F_b\rangle=|F_{n_3+3*x}\rangle$. If Bob has $|F_{n_3+3*x}\rangle$, he will only be able to dope out one of Alice's value through her first bit if she sends $B_{n_3+x}$. If he has $|F_{n_3}\rangle$ , he will only be able to know any one of Alice's value through her first bit if she sends $B_{n_3+2*x}$. So Bob sending the recognition flag  would be sufficient for Alice to deduce what OAM value Bob has.

Because Bob's bit is a central bit, its minimum distance to the nearest opposite bit should be $2$ (The distance between $F_n$ and $F_m$ is defined as $|n-m|$), and the nearest opposite bit is $B_{n_3+2*x}$. So Bob surmises that Alice should own $|F_{n_3+2*x}\rangle$.

Also because all three OAM values Alice and Bob share should be three adjacent third-order Fiboncci numbers, and because Bob already has $|F_{n_3}\rangle$, he can deduce that Alice should also own $|F_{n_3+x}\rangle$ which is between $|F_{n_3+2*x}\rangle$ and $|F_{n_3}\rangle$.

 So after receiving Bob's response bit, Alice would send $B_{n_3+2*x}$ which is equal to $B_{n_3}$.

\item
Alice sends her bit according to her first rule.

So she has $|F_{n_3+x}\rangle \otimes |F_{n_3-x}\rangle$ and she sends $!B_{n_3+x}$ and $!B_{n_3-x}$ respectively to Bob regardless of the order while $!B_{n_3+x}=!B_{n_3-x}\not=B_b=B_{n_3}$. In this situation, she already knows that Bob's value is  $F_{n_3}$. So what bit Bob would send doesn't actually matter here for it won't affect Alice's correct guess of Bob's OAM value.

It is worth mentioning that the second rule would not be confused with the first rule as there is only one opposite bit within the maximum distance to $B_{n_3}$ to maintain that all three OAM values are adjacent: $B_{n_3+2*x}$.
So when two opposite bits are sent to Bob, this contradiction would eliminate the possibility of the second rule. During the communication under Alice's second rule, Bob would receive "01" while under her second rule he would receive "00", which would be sufficient for him to distinguish the two cases.

\end{enumerate}

In both cases above Alice should have $|F_{n_3+x}\rangle$. So Bob should send Alice "$1$" to illustrate that he is able to identify one of Alice's value.


\item If Alice sends Bob $B_{a_1}=B_b=B_{n_3}$, then she can only be sending it according to her first rule.

So Alice could have sent any one in $\{B_{n_3+x}, B_{n_3-x}, B_{n_3-2*x}\}$
and she could have $|F_{n_3+2*x}\rangle\otimes |F_{n_3+x}\rangle$ or $|F_{n_3-x}\rangle\otimes |F_{n_3-2*x}\rangle$
. As the two possibilities share no common value of Alice, Bob should send "$0$" to illustrate his inability to acknowledge any of Alice's bits.

\begin{enumerate}
\item
If Alice has $|F_{n_3+2*x}\rangle\otimes |F_{n_3+x}\rangle$
, she would surmise that Bob has $|F_{n_3+3*x}\rangle$ or $|F_{n_3}\rangle$
, each requires a different bit from Alice in order to identify one of Alice's values. This is the same as 1.(a) and Bob would be able to acknowledge all of Alice's values after receiving all two bits from her.

\item
If Alice has $|F_{n_3-x}\rangle\otimes |F_{n_3-2*x}\rangle$
, she would surmise that Bob has $|F_{n_3-3*x}\rangle$ or $|F_{n_3}\rangle$.
 If Bob has $|F_{n_3-3*x}\rangle$
, he would be able to identify one of Alice's bit through the first bit from her. He wouldn't if he has $|F_{n_3}\rangle$
.
\end{enumerate}

So sending the recognition flag here can still help Alice  know Bob's value. After that, Alice would send the other bit she has to Bob, which should be able to make Bob fully aware of Alice's values.

\end{enumerate}


Bob has the photon with OAM value $F_b=F_{n_3}$ while $(n_3$ mod $4)\in\{0, 1\}$. (Bob's bit is an "edge" bit.)
\begin{enumerate}
\item
If Alice sends Bob $B_a = B_b = B_{n_3}$, then Bob will know that $B_a\in\{B_{n_3+x}, B_{n_3+2*x}\}$. Alice should possess  $|F_{n_3+x}\rangle\otimes |F_{n_3+2*x}\rangle$ or $|F_{n_3+x}\rangle\otimes |F_{n_3-x}\rangle $, with a common value $|F_{n_3+x}\rangle$. So Bob sends "1".

\begin{enumerate}
\item
If Alice has $|F_{n_3+x}\rangle\otimes |F_{n_3-x}\rangle $ , then she knows that Bob's value is  $F_{n_3}$. So what bit Bob would send doesn't actually matter here for it won't affect Alice's correct guess of Bob's OAM value.

After Bob sends his bit to Alice, Alice would send another bit of hers to Bob, which is opposite to the first bit she sent to Bob. And because Bob's bit is surrounded closely by two different bits, this would be enough for Bob to know the exact bits Alice has.

\item
If Alice has $|F_{n_3+x}\rangle\otimes |F_{n_3+2*x}\rangle$, she would only surmise that Bob has $|F_{n_3}\rangle$ or $|F_{n_3+3*x}\rangle$. Because those two bits are of the same category (they are both "edge" bits), their only difference is whether they are on the "left" or the "right" side of their set. We stipulate that if Bob's bit ($B_b$) is on the "left" edge of its set, or in another way, $B_{b-1}\not=B_{b}$ and $B_{b+1}= B_{b}$, then Bob would send "$1$" to Alice, otherwise if $B_{b-1}= B_{b}$ and $B_{b+1}\not=  B_{b}$, then Bob would send "$0$" to Alice, which forms Bob's second rule to send$!(n_3\text{ mod }4)$. This would be sufficient for Alice to acknowledge Bob's value.

After this, Alice would send another bit of hers to Bob, leading to their full understanding of each other's values.
\end{enumerate}

\item If Alice sends Bob $B_a \not= B_b = B_{n_3}$, she should possess $|F_{n_3+x}\rangle\otimes |F_{n_3-x}\rangle$ or $|F_{n_3-2*x}\rangle\otimes |F_{n_3-x}\rangle $.

\begin{enumerate}
\item
If Alice has $|F_{n_3+x}\rangle\otimes |F_{n_3-x}\rangle $, Alice would know Bob's value needless of his bit. And she would be able to inform him of his value later with another bit like in 1.(a).

\item
If Alice has  $|F_{n_3-2*x}\rangle\otimes |F_{n_3-x}\rangle $, she should surmise that Bob has $|F_{n_3-3*x}\rangle$ or $|F_{n_3}\rangle$. Because if Bob has $|F_{n_3-3*x}\rangle$ , he can't determine what value Alice has (let's assume that $B_0=1$ to accord the third-order Fibonacci recursion) while he can if he has $|F_{n_3}\rangle$. So Bob could send the recognition flag to Alice as is stated in Bob's rule.

After that Alice would send another bit of hers to pass on all the information Bob needs to know Alice's values.
\end{enumerate}
\end{enumerate}

After these process, Alice and Bob can add up their values to get the pump value, which serves as one segment of the key.

\subsection{An example of the communication process}

Here we give an example of the communication process.

If Alice measures her OAM values as 11 and 20, then Alice can know that Bob has 6 or 37 (see \ref{t1}). According to Alice's rules, she now has two adjacent values in Fibonacci sequence (the values 11 and 20 are adjacent). So she sends the allocated bit for either OAM value 11 or 20 and sends another one after receiving Bob's bit (according to \ref{alice}).

Let's say that she sends the bit for OAM value 11, which is 1 in \ref{t1}. Now, if Bob has OAM value 6, he would assume that Alice has OAM values as "2+3" or "3+11" or "11+20". So now with the first bit "1" from Alice, he can know that the combination "2+3" is not possible. He still needs another bit from Alice to determine the exact values she has. But before that, he sends his bit according to \ref{bob}, which is 1.

Now Alice receives Bob's "1". So she knows that Bob has the OAM vales "6" because the bit allocated for OAM value 37 is "0" according to \ref{bob}. Now Alice send the bit allocated for the other OAM value of hers (20), which is "1", to let Bob know her exact OAM values.

So now Bob has the second bit from Alice, "1". This gives him two "1"s so the possibility of the combination "3+11" is eliminated. Now Bob knows that Alice has OAM values "11+20".
And with Alice knowing Bob's OAM value, both communication parties now know all three OAM values distributed between them and they can generate secret keys based on preappointed rules.

Similarly, we can analyze the above case in which Bob has the OAM value "37". The protocol can also work with other combinations Alice may have. Please notice that although we allocated a bit for every OAM values in \ref{t1}, the bits are not directly used in classical channel. The sending strategy for Alice and Bob are listed in \ref{alice} and \ref{bob}. And usually what Bob is sending doesn't has a lot to do with the original bit allocated for his OAM value (\ref{bob}). But with this complex protocol well-known between the two parties, we are able to increase the information capacity and make sure both sides are able to get the right information following the protocol.


\section{The security performance of the QKD protocol}

Traditionally, the security checking of the protocol is based on the polarization degrees of freedom. In the proposed QKD protocol, by randomly choosing the photons in the checking mode, the two communication parties could discover the eavesdropping behavior \cite{r5}. Moreover, the security checking process could be realized by using the OAM degrees of freedom. During the detection of eveasdropping, if Alice measures $\{l_{a_1}, l_{a_2}\} = \{6, 20\}$, the state reaches Bob should be $|{11}\rangle$; if Alice measures $\{l_{a_1}, l_{a_2}\} = \{3, 6\}$ or $\{l_{a_1}, l_{a_2}\} = \{20, 37\}$, the state reaches Bob should be $\frac{1}{\sqrt{2}}\{|{2}\rangle+|{11}\rangle\}$ or $\frac{1}{\sqrt{2}}\{|{11}\rangle+|{68}\rangle\}$. Suppose there is an eavesdropper in the channel, called Eve, who perform the intercept-resend attack on the protocol. For example, she intercepts Bob's photon and reads out the value $l_b = 11$. However, she could not able to know what photon to resend to Bob: $|{11}\rangle$, $\frac{1}{\sqrt{2}}\{|{2}\rangle+|{11}\rangle\}$ or $\frac{1}{\sqrt{2}}\{|{11}\rangle+|{68}\rangle\}$. The wrong photons would change the right distribution of each value in Bob's measurement, which would expose her eveasdropping. And it's worth mentioning that as shown in TABLE \ref{t2}, the classical information exchanged between Alice and Bob is insufficient for an eavesdropper to determine the value.

\newcommand{\tabincell}[2]{\begin{tabular}{@{}#1@{}}#2\end{tabular}}

\begin{table*}[!hbp]
\centering
\begin{tabular}{|c|c|c|c|c|c|c|c|c|}
\hline
Eve sees & 000 & 001 & 010 & 011 & 100 & 101 & 110 & 111 \\
\hline
Secret key & 6,11 & 20,423 & \tabincell{c}{11,20,68,\\125,423} & 20,37,230 & 37,230 & 68,125 & \tabincell{c}{20,37,230,\\423} & \tabincell{c}{6,11,37,\\125,230} \\
\hline
\end{tabular}
\caption{In the top row, the first and the third digit in each pair are the classical bits sent by Alice separately, the second digit is the bit sent by Bob. The second row shows all corresponding possible secret keys.}
\label{t2}
\end{table*}
Assuming an intercept-resend attack is performed by Eve, the average probability of a correct guess is $37.92\%$. The probability would drop according to the increment of the number of Fibonacci values used.
\begin{table*}[!hbp]
\centering
\begin{tabular}{|c|c|c|c|c|c|c|c|c|}
\hline
Eve sees & 000 & 001 & 010 & 011 & 100 & 101 & 110 & 111 \\
\hline
Secret key & \tabincell{c}{3,5,355\\ } &  \tabincell{c}{3,9,355\\ } & \tabincell{c}{9,17,26,\\ 162,193} & \tabincell{c}{9,17,26,\\ 162,193}  & 31,105 & 31,105,57 & \tabincell{c}{2,14,17,\\31,105,\\193,298} & \tabincell{c}{2,14,17,\\31,105,\\193,298} \\
\hline
\end{tabular}
\caption{In the top row, the first and the third digit in each pair are the classical bits sent by Alice separately, the second digit is the bit sent by Bob. The second row shows all corresponding possible secret keys.}
\label{t3}
\end{table*}
\begin{table*}[!hbp]
\centering
\begin{tabular}{|c|c|c|c|c|c|c|c|c|}
\hline
Eve sees & 000 & 001 & 010 & 011 & 100 & 101 & 110 & 111 \\
\hline
Secret key & \tabincell{c}{3,6,11\\ } &  \tabincell{c}{2,20,\\423,\\ 28750 } & \tabincell{c}{2,14,51,17,\\31,105,193,\\298,13125,\\44390} & \tabincell{c}{0,33,30,\\ 66,2516,\\15640,\\8500}  & \tabincell{c}{9,17,26,\\ 162,193\\1147,\\3876} & \tabincell{c}{18,66,99,\\120,340,\\4625,\\7141,\\8500} & \tabincell{c}{37,31,\\230\\ } & \tabincell{c}{220,68,\\125} \\
\hline
\end{tabular}
\caption{In the top row, the first and the third digit in each pair are the classical bits sent by Alice separately, the second digit is the bit sent by Bob. The second row shows all corresponding possible secret keys.}
\label{t4}
\end{table*}
Also the protocol is intrinsically immune to photon-number splitting attack. If a multi-photon pulse is sent into the spiral source, there is no reason for the photons that come out to be in the same state: they are distributed among the different third-order Fibonacci numbers in the same manner as they would if they had been sent one by one. Siphoning off one photon from a pulse will reveal nothing about the state of the other photons in that pulse to the eavesdropper, thus making the attack invalid for Eve.

Moreover, the security checking could be improved by simply changing some unnecessary bits of Alice into the random bits. From the procedure analysis of the protocol, we know that under certain circumstances, Bob would be able to acknowledge at least one value of Alice's through only the first bit of hers. If we change this bit into a random bit or to indicate the method of calculating secret keys, to add up or to multiply for example, the ambiguity will be increased, making it harder for Eve to guess the right value from the classical information exchange. Meanwhile we can make simple rules to determine how to process the values when Bob knows all of them to get the secret keys.

Here we simply change that bit into the random bit. For every case in which Bob would be able to know all Alice's values, we stipulate they choose the smallest two values to add up to get the secret key. The result is shown in TABLE \ref{t3}. For every case in which Bob would be able to know all Alice's values, we assume that they choose the smallest two values to add up to get the secret key. The result is shown in TABLE \ref{t3}.

This table has shown the first step in improving the protocol, and Eve's average correct guessing rate will be decreased to $27.32\%$.

Some values, like "$3$", "$355$", are actually calculated in several different cases(000, 001). This is because in every case in which Bob can acknowledge all Alice's values, we stipulate that the smallest two values be added to get the secret key.  By choosing the values with common rules to avoid as many overlap as possible, further improvement can be achieved. Here if we stipulate that they choose digits and the calculating method (add or multiply) differently according to $(n_3 \text{ mod } 4)$ ($l_b=l_{n_3}$), the average correct guess rate can be decreased to $22.01\%$, as shown in TABLE \ref{t4}.

Notice that the distribution of possibilities regarding every classical information exchange case is not balanced. So we can only retain the cases with a relatively better performance (with more possible secret key values) to further improve the protocol. As an extreme case, we only retain the results of the classical information exchange "$010$" (the case with the most possible key values) and discarded all the others, we may achieve a minimum Eve's correct guessing rate as $10\%$.

For a coding space of $N$ Fibonacci values, this rate can be decreased to $\frac{9}{10*(N+1)}$, with information entropy density increased to   $\log_{2} \frac{10*(N+1)}{9} $ per photion. Here we simulated the performance of the protocol in Fig.2 and Fig.3, with the method shown in \cite{Rnet}.

First in Fig.2, the relation between the keys rate and the communication distance is presented. Here the frequency $f_{rep}$ is set as 10MHz, the intensity of pulse is 0.1 photon per pulse and the channel loss of the fiber is 0.2dB/km. The efficiency of the detector is set as 0.1, and the dark count of the detector is $10^{-4}$ per second. The imperfect interference or polarization contrast induced quantum bit error rate is neglected. We found that the keys generation rate increases with the increment of coding space of the OAM states.
\begin{figure}
\centering
\includegraphics[width=3in]{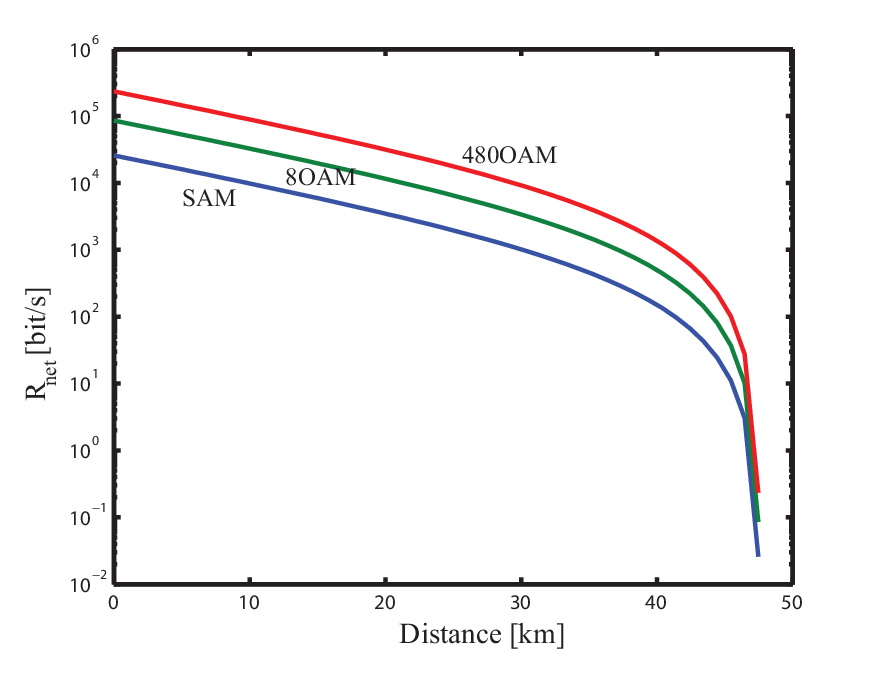}
\caption{Bit rate, after error correction and privacy amplification, vs fiber length for QKD protocols with respective coding space of SAM, 8 OAM states and 480 OAM states.}
\label{RD}
\end{figure}
In Fig.3, we present the relation of the coding bits and security performance with different coding space. In Fig.3(a), the bits values of each photon is enlarged by using our protocol compared with traditional method without Fibonacci\cite{firstoamqkd}. And in Fig.3(b), the security of the protocol is improved as the eavesdropping behavior will easily be discovered.
\begin{figure}
\centering
\includegraphics[width=4in]{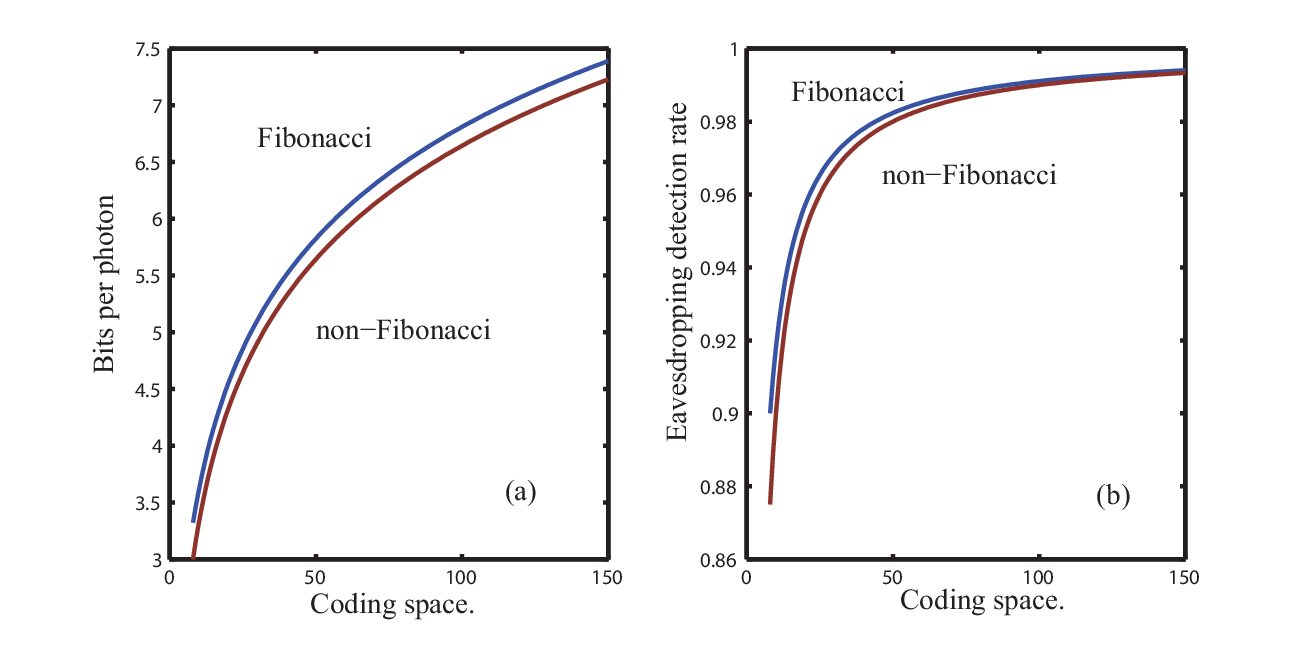}
\caption{Figure (a), (b) shows that information entropy carried by every photon and eavesdropping detection rate increase with the enlargement of coding space. Both figures give the performance of both Fibonacci and non-Fibonacci OAM protocols.}
\label{BEC}
\end{figure}

\section{Summary}

In summary, exploiting the OAM states based on third-order Fibonacci recursion, we proposed a high-capacity encoding scheme for quantum key distribution. The secure key distribution is realized by using three particles entangled state and third-order Fibonacci sequence which could be generalized to high-order sequence and multi-parties quantum network. Compared with the traditional method, the probability for detecting the eavesdropper could be increased to $90\%$ with only $N=8$ OAM values, which can be further improved with the increment of the number of OAM values used. Also the protocol is intrinsically immune to photon splitting attack and the security can be further enhanced implementing a new type of decoy state.

%

\bibliographystyle{IEEEtran}
\bibliography{QSDC}






\end{document}